\documentstyle[preprint,aps,tighten,epsfig]{revtex}

\begin{document}
 
\title{A quark-model based study of the triton binding energy}
 
\author{B. Juli\'a-D\'\i az $^{(1)}$, J. Haidenbauer $^{(2)}$, 
A. Valcarce $^{(1,3)}$, and F. Fern\'andez $^{(1)}$ }
\address{$(1)$ Grupo de F\' \i sica Nuclear,
Universidad de Salamanca, E-37008 Salamanca, Spain}
\address{$(2)$ Institut f\"ur Kernphysik (Theorie),
Forschungszentrum J\"ulich, D-52425 J\"ulich, Germany}
\address{$(3)$ Departamento de F\' \i sica Te\'orica, 
Universidad de Valencia, E-46100 Valencia, Spain}
\maketitle

\begin{abstract}
The three-nucleon bound state problem is studied
employing a nucleon-nucleon potential obtained
from a basic quark-quark interaction 
in a five-channel Faddeev calculation. The obtained
triton binding energy is comparable to those predicted 
by conventional models of the $NN$ force. 

\vspace*{2cm} \noindent Keywords: \newline
non-relativistic quark models, triton binding energy, Faddeev calculations \newline
\newline
\noindent Pacs: \newline
 12.39.Jh, 21.45.+v, 13.75.Cs, 21.10.Dr 
\end{abstract}

\newpage    

\section{Introduction}
During the last decade the development of quark-model 
based interactions for the hadronic force
has led to nucleon-nucleon ($NN$) potentials that provide 
a fairly reliable description of the on-shell data. 
Several models including quark-degrees of freedom have
been used to study the $NN$ interaction \cite{SHIMIZU} and also the
baryon spectra \cite{ISGUR}. Among them, the chiral quark cluster
model is the only one that pursued a simultaneous 
understanding of different low-energy phenomena
based on a unique quark-quark interaction. This model
is able to provide a quantitative description 
of the $NN$ scattering \cite{jpg} and bound state
problems \cite{buchmann}. 

Nevertheless, quark-model based $NN$ interactions have not been often used 
to study few-body systems. There might be two different reasons for that.
First of all, most of those interaction models for the
two-nucleon system needed to be supplemented with 
meson-exchange potentials between the baryons to obtain 
a reasonable description of the experimental data \cite{Oka,Tuebingen},
loosing in this way their quark-based character. Secondly, other quark-model
based interactions were primarily designed to describe the baryon 
spectra \cite{riska}, but lead to unrealistic results when they are applied 
to the two-nucleon system \cite{nakamoto}.

In this work we want to perform a study of the 
triton bound state making use of a nonlocal $NN$ potential fully 
derived from quark-quark interactions. The model has been 
previously utilized for investigations of three-body systems 
($NNN$, $NN\Delta$, $N\Delta\Delta$, and $\Delta\Delta\Delta$),
making more emphasis on the mass ordering of possible 
bound states of these systems than on the binding
energy values\cite{hum}. In the
present work the full nonlocal $NN$ potential will be employed
as it follows from the application of the resonating group method 
(RGM) formalism. This method allows, once the Hilbert space for the 
six-body problem has been fixed, to treat the 
inter-cluster dynamics in an exact way. Thereby, 
nonlocalities are generated, reflecting the internal structure of the 
nucleon, which translate into specific off-shell properties 
of the resulting $NN$ potential. 

Indeed the relevance and/or necessity of considering the nonlocal parts 
of $NN$ potentials in realistic interactions is still under debate. 
Over the past few years several studies have appeared
in the literature which stress the potential importance
of nonlocal effects for the 
quantitative understanding of few-body observables and,
specifically, for the triton binding energy 
\cite{TA92,Gibson,HH,MSS,Elster,DO00}. However, 
the majority of these investigations \cite{Gibson,HH,MSS,Elster,DO00}
explore only nonlocalities arising from the meson-exchange
picture of the $NN$ interaction.

The nonlocalities
generated in a quark-model derivation of baryonic potentials
may play a relevant role for the case of 
the three-nucleon bound state. It has been
argued that the assumptions associated with meson-exchange
models sharply limit the nature of the off-shell
properties of those potentials, once the on-shell matrix
elements are constrained to fit the $NN$ data \cite{stricker}.
Therefore, it is very interesting to investigate the off-shell
features of potentials derived from a quark-model. Some
preliminary studies in this direction have been done in Ref. \cite{TA92}.
However, there only the short-range part of the interaction is obtained
by means of quark-model techniques. The intermediate and long-range part is
described by ``standard'' meson-exchange between baryons.
Accordingly, that work allows only very limited conclusions with regard
to effects of the quark substructure.

The triton binding energy is obtained from a Faddeev calculation. 
We restrict ourselves to the standard five channel case; i.e. we 
consider only the $^1S_0$ and $^3S_1-{}^3D_1$ $NN$ partial waves, 
those which provide the bulk contribution to the 
three-nucleon binding energy. The three-body Faddeev equations
will be solved in momentum space making use of separable finite-rank
expansions of the two-body interactions.

The paper is organized as follows. In Section II we introduce
the basic quark-quark Hamiltonian and we describe  
the method to obtain the RGM $NN$ interaction.
In Section III we provide details about the
finite-rank expansions of the quark-model based potentials
which enter in the Faddeev calculations of the triton binding energy
and we present results for the three-body system. 
Finally, some concluding remarks are provided in Sec. IV.

\section{Quark-model based $NN$ potential}
In recent years a chiral quark cluster model for the $NN$
interaction has been developed. This model has been widely 
described in the literature \cite{jpg,buchmann,AVA95,DEN1}, therefore
we will only briefly summarize here its most relevant aspects.
It contains, as a consequence of 
chiral symmetry breaking, a pseudoscalar and a scalar exchange
between constituent quarks coming from the Lagrangian
\begin{equation}
L_{ch}=g_{ch} F(q^2)\bar{\Psi}(\sigma + i\gamma_5 \vec{\tau}\cdot\vec{\pi})\Psi \, ,
\end{equation}
where $F(q^2)$ is a monopole form factor
\begin{equation}
F(q^2) =\left [  \Lambda_{\chi}^2 \over \Lambda_{\chi}^2 +q^2 \right ]^{1\over2} \,
.
\end{equation}
$\Lambda_{\chi}$ determines the scale of chiral symmetry breaking,
being bound between 1 GeV and 600 MeV \cite{henley}. The chiral coupling
constant $g_{ch}$ is chosen to reproduce the experimental $\pi NN$ coupling
constant.

>From the above Lagrangian a pseudoscalar ($PS$) and a scalar ($S$) potential
between quarks can be easily derived in the non-relativistic 
approximation,
\begin{equation} 
V_{ij}^{PS}(\vec{q}) = - {g_{ch}^2 \over 4 m_q^2 }
{\Lambda_{\chi}^2 \over \Lambda_{\chi}^2 +q^2 } 
{(\vec{\sigma}_i\cdot\vec{q})(\vec{\sigma}_j\cdot\vec{q}) 
\over m_{PS}^2+ q^2 } (\vec{\tau}_i\cdot\vec{\tau}_j) ,
\end{equation}
\begin{equation} 
V_{ij}^{S}(\vec{q}) = -g_{ch}^2 
{\Lambda_{\chi}^2 \over \Lambda_{\chi}^2 +q^2 }
{1\over m_{S}^2+ q^2 } \, .
\end{equation}

Using the range of values for $\Lambda_{\chi}$ given above
yields a $N-\Delta$ mass difference due to the pseudoscalar interaction 
between 150 and 200 MeV.
The rest of the mass difference, up to the experimental value, must
have its origin on perturbative processes. In the present model, this is taken into 
account through the one-gluon exchange potential 
\cite{RUJ},
\begin{equation}
V_{ij}^{OGE}(\vec{q})  = \alpha_s (\vec{\lambda}_i\cdot
 \vec{\lambda}_j)
  \left \{ {\pi \over q^2} - {\pi\over 4 m_q^2} \left [
1 + {2\over3} (\vec{\sigma}_i\cdot\vec{\sigma}_j) \right ] 
  + {\pi\over 4 m_q^2} {{[\vec{q}\otimes\vec{q}]^{(2)}\cdot
[\vec{\sigma}_i\otimes\vec{\sigma}_j]^{(2)}} \over q^2}  \right \},
\end{equation}
where the $\lambda$'s are the color Gell-Mann matrices and 
$\alpha_s$ is the strong coupling constant.

For the present study we make use of the nonlocal $NN$ potential
derived through a Lippmann-Schwinger formulation of the
RGM equations in momentum space \cite{DEN1}.
The formulation of the RGM for a system of two 
baryons, $B_1$ and $B_2$, needs the wave function
of the two-baryon system constructed from 
the one-baryon wave functions. The two-baryon
wave function can be written as:

\begin{equation}
\Psi_{B_1 B_2}={\cal A} [\chi(\vec{P}) \Psi_{B_1 B_2}^{ST} ]= 
{\cal A} [\chi(\vec{P}) \phi_{B_1}(\vec{p}_{\xi_{B_1}}) 
\phi_{B_2}(\vec{p}_{\xi_{B_2}}) 
\chi_{B_1 B_2}^{ST} \xi_c [2^3]] ,
\end{equation}

\noindent
where ${\cal A}$ is the antisymmetrizer of the six-quark system,
$\chi(\vec{P})$ is the relative wave-function of the two clusters,
$\phi_{B_1}(\vec{p}_{\xi_{B_1}})$ is the internal spatial 
wave function of the baryon $B_1$, $\xi_{B_1}$ are the internal 
coordinates of the three quarks of baryon $B_1$. $\chi_{B_1 B_2}^{ST}$
denotes spin-isospin wave function of the two-baryon system coupled
to spin ($S$) and isospin ($T$), and, finally, $\xi_c [2^3]$
is the product of two color singlets.

The dynamics of the system is governed by the 
Schr\"odinger equation
\begin{equation}
({\cal H} -E_T)| \Psi>=0 \ \ \ \Rightarrow \ \ \ <\delta \Psi | ({\cal H} -E_T)
| \Psi>=0, 
\label{variations}
\end{equation}
where 
\begin{equation}
{\cal H} = \sum_{i=1}^N {\vec{p_i}^2 \over 2 m_q} +\sum_{i<j} V_{ij}
-T_{c.m.}
\end{equation}
with $T_{c.m.}$ being the center of mass kinetic energy, $V_{ij}$
the quark-quark interaction described above, and $m_q$
the constituent quark mass.

Assuming the functional form 

\begin{equation}
\phi_B( \vec p ) = {\left( {b^2 \over \pi} \right)}^{3/4} e^{{-b^2 p^2} \over 2} \,
,
\end{equation}

\noindent
where $b$ is related to the size of the nucleon quark core,
Eq. (\ref{variations}) can be written in the following way,
after the integration of the internal cluster degrees of freedom,

\begin{equation}
\left ( {\vec{P}^2 \over 2\mu} -E \right ) \chi(\vec{P}) +
\int \left( V_D (\vec{P}, \vec{P}_i) +
W_L (\vec{P}, \vec{P}_i) \right)
\chi(\vec{P}) d\vec{P}_i =0
\label{ALS}
\end{equation}

\noindent
$V_D (\vec{P}, \vec{P}_i)$ is the direct RGM kernel and 
$W_L (\vec{P}, \vec{P}_i)$ is the exchange RGM kernel, composed
of three different terms

\begin{equation}
W_L (\vec{P}, \vec{P}_i) = T_L (\vec{P}, \vec{P}_i) + 
V_L (\vec{P}, \vec{P}_i) + (E + E_{in}) K_L (\vec{P}, \vec{P}_i) 
\, , \label{eq11}
\end{equation}

\noindent
where $E_{in}$ is the internal energy of the two-body system,
$T_L (\vec{P}, \vec{P}_i)$ is the kinetic energy exchange 
kernel, $V_L (\vec{P}, \vec{P}_i)$ is the potential 
energy exchange
kernel and $K_L (\vec{P}, \vec{P}_i)$ is the exchange norm 
kernel. Note that if we do not mind how 
$V_D (\vec{P}, \vec{P}_i)$ and 
$W_L (\vec{P}, \vec{P}_i)$ were derived microscopically,
Eq. (\ref{ALS}) can be regarded as a general single channel equation
of motion with including energy-dependent non-local potential.
$V_D(\vec{P}, \vec{P}_i)$, which
contains the direct RGM potential, and 
$W_L(\vec{P}', \vec{P}_i)$, which
contains the exchange 
RGM potential coming from
quark antisymmetry, constitute our 
energy-dependent nonlocal potential. 
In our case $E_{in}= 2 m_N$ what makes
our potential almost energy-independent, because the center
of mass energy of the two-body system $E$ is much smaller
than the internal energy $E_{in}$. 

The potential yields a fairly good reproduction
of the experimental data up to laboratory energies of 250 MeV. 
For a correct description of the $^1S_0$ phase
shift it is necessary to take into account the coupling
to the $^5D_0$ $N\Delta$ channel \cite{AVA95}, that 
provides an isospin-dependent mechanism generating the additional
attraction in this channel. This is implemented in our calculation
generalizing Eq. (\ref{ALS}) to a coupled channel scheme. 
It implies a modification of Eq. (\ref{eq11}) with an additional
term which contains the $NN \to N\Delta$ coupling. The parameters used are
summarized in Table \ref{table1}. In Table \ref{table2} we present
the results for the low-energy scattering data and the deuteron properties 
of the present model together with values of 
some standard $NN$ potentials \cite{Mach,nijmegen} and experimental data. 
It is known that a charge symmetry breaking term should be included 
in the interaction if one wants to reproduce those quantities
simultaneously \cite{ENTE}. This is taken into account by a slight
modification of the chiral coupling constant
to reproduce the
deuteron and the low energy scattering parameters (see Table \ref{table1}).
We also show, in Figs. \ref{fig1}, \ref{fig2} and \ref{fig3}, the $^1S_0$ and 
$^3S_1-{}^3D_1$ phase shifts and the mixing parameter $\varepsilon_1$
in comparison to results from phase-shift analyses \cite{ex1,ex2,ex3}.
 
\section{Triton binding energy}
The triton binding energy is obtained by means of a 
Faddeev calculation using the $NN$ interaction described 
above.
We perform a so-called five-channel calculation, 
i.e., we use only the $^1S_0$ and $^3S_1-{^3D_1}$ $NN$ 
partial waves as input. Note that since in our model there 
is a coupling to the $N\Delta$ system, as explained above,
a fully consistent calculation would require the inclusion of 
two more three-body channels. However, their
contribution to the $3N$ binding energy is known to be rather small \cite{Hajd}
and therefore we neglect them for 
simplicity reasons. 

To solve the three-body Faddeev equations in momentum space
we first perform a separable finite-rank expansion of the
$NN(-N\Delta)$ sector utilizing the EST method \cite{EST}. 
Such a technique has been extensively studied  
by one of the authors (J.H.) 
for various realistic $NN$ potentials \cite{WSH00} and
specifically for a model that also includes a coupling to the 
$N\Delta$ system \cite{NEMO}. In those works it was shown that,
with a separable expansion of sufficiently high rank, 
reliable and
accurate results on the three-body level can be achieved. 
In the present case it turned out that separable 
representations of rank 6 -- for $^1S_0-(^5D_0)$ and for
$^3S_1-{^3D_1}$ -- are sufficient to get converged 
results. The set of energies used for the EST separable 
representations are listed in Table \ref{table3}. We refer
the reader to Refs.~\cite{WSH00,NEMO,Haid84} for 
technical details on the expansion method.
The quality of the separable expansion on the $NN$ sector 
can be seen in Fig. \ref{fig1}, where we show phase shifts 
for the original nonlocal potential and for the corresponding separable 
expansion. Evidently, the phases are almost indistinguishable.

Results for the triton are summarized in Table IV. 
It is reassuring to see that the predicted triton binding energy 
is comparable to those obtained from conventional $NN$
potentials, such as the Bonn or Nijmegen models. 
Thus, our calculations show that quark model based $NN$ interactions are 
definitely able to provide a realistic description of the triton. The
results also give support to the use of such interaction model for
further few-body calculations.
One should not forget at this point that the number of free parameters
is greatly reduced in quark-model based $NN$ interactions like 
the present one. 
In addition, the parameters are strongly correlated by the requirement to 
obtain a reasonable description of the baryon spectrum.

\section{Conclusions}
We have calculated the three-nucleon bound state problem 
utilizing a nonlocal $NN$ 
potential derived from a basic quark-quark interaction. 
This potential was generated by means of the
resonating group method so that nonlocalities resulting
from the internal structure of the nucleon were preserved.
The resulting triton binding energy is comparable to those
obtained from conventional $NN$ potentials. 

In the calculation of the three-nucleon binding energy we
have followed the traditional approach, namely solving the
Faddeev equations with nucleon degrees of freedom. 
Let us remark, however, that in a more fundamental approach 
one would impose consistency between the treatment of the two-
and three-nucleon systems in terms of quark degrees of
freedom. That, of course, would require the derivation and
solution of corresponding three-nucleon RGM equations. 
In such a framework the quark structure of nucleons generates 
(besides the consecutive two-nucleon interactions that are summed up 
by the Faddeev equations) also genuine three-body forces. 
These forces are of short-ranged nature and they could be 
significant for short-distance phenomena like the
high-momentum-transfer part of the charge form factor of $^3$He. 
Indeed, there have been attempts to explore the effects
of such three-body forces on the triton binding energy. In a
simple model based on a single one-gluon exchange \cite{HECHT}
the three-body exchange kernels have been evaluated.
An estimation provided in this reference suggests that those 
three-nucleon forces could yield additional binding in the order 
of 0.2 MeV. If this is the case then those effects would be
still small enough to guarantee that the approach we followed in our 
study is sufficiently accurate for an exploratory calculation.
However, one has to keep in mind that the estimation in Ref.~\cite{HECHT}
was done only in perturbation theory, and by means of a zeroth order 
three-nucleon wave function with a series of fitted parameters. 
Thus, for the future, a more refined and consistent treatment 
of the three-nucleon problem within the quark picture is certainly
desirable in order to allow for reliable conclusions on this issue. 

\acknowledgements
The authors thank D.R. Entem for providing the 
codes used for the nonlocal interaction potential and for
many useful comments. One of the authors (B.J-D.) wants to thank the
hospitality of the FZ J\"ulich where part of this work has been done.
A.V. thanks the Ministerio de Educaci\'on, Cultura y Deporte of Spain
for financial support through the Salvador de Madariaga program.
This work has been partially funded by Direcci\'{o}n General de
Investigaci\'{o}n Cient\'{\i}fica y T\'{e}cnica (DGICYT) under the Contract
number PB97-1401, and by Junta de Castilla y Le\'on under the Contract number
SA-109/01.

\begin{table}[h]
\caption{Quark model parameters. The values in brackets are used for 
a correct description of the deuteron.}
\begin{center}
\begin{tabular}{cccc}
& $m_q ({\rm MeV})$          & 313     &       \\ 
& $b
\tablenote{\small{$b$ is the parameter of the harmonic oscillator               
wave function used for each quark $\eta(x) = \left({1 \over                     
{\pi b^2}} \right)^{(3/4)} \,\, e^{-(x^2/2b^2)}$.}}  
({\rm fm})$             & 0.518     &     \\ 
& $\alpha_s$              & 0.4977  &     \\ 
& $g_{ch}^2$              & 6.60(6.86)  &        \\ 
& $m_S ({\rm fm^{-1}})$  &3.400  &  \\ 
& $m_{PS} ({\rm fm^{-1}})$    & 0.70   &       \\ 
& $\Lambda_{\chi} ({\rm fm^{-1}})$   & 4.47 &    \\
\\ \end{tabular} \end{center}
\label{table1}
\end{table}

\begin{table}[h]
\caption{$NN$ properties}
\label{table2}
\begin{tabular}{llllll}
& & Quark model & Nijm II \protect\cite{nijmegen}& 
Bonn B\protect\cite{Mach} &Exp. \\
\tableline 
\tableline
\multicolumn{6}{c}{ Low-energy scattering parameters} \\
\tableline
 $^1S_0$ & $a_s$ (fm) & $-$23.759  &-23.739& $-$23.750&
$-$23.74$\pm$0.02  \\
         & $r_s$ (fm) &  2.68     &2.67& 2.71&2.77$\pm$0.05\\
 $^3S_1$ & $a_t$ (fm) & 5.461      &5.418 &5.424&5.419$\pm$0.007 \\
         & $r_t$ (fm) & 1.820      &1.753 & 1.761&1.753$\pm$0.008\\
\tableline
\tableline
\multicolumn{6}{c}{ Deuteron properties} \\
\tableline
\multicolumn{2}{l}{$\epsilon_d$ (MeV)} &$-$2.2242  &
$-$2.2246&$-$2.2246&$-$2.224575  \\
\multicolumn{2}{l}{$P_D$ ($\%$)}   & 4.85   & 5.64  & 4.99  &$-$\\
\multicolumn{2}{l}{$Q_d$ (fm$^2$)} &0.276  & 0.271  & 0.278 &
0.2859$\pm$0.0003\\
\multicolumn{2}{l}{$A_S$ (fm$^{-1/2}$)}& 0.891   & 0.8845 &
0.8860&0.8846$\pm$0.0009\\
\multicolumn{2}{l}{$A_D / A_S $}      &0.0257    & 0.0252 &
0.0264&0.0256$\pm$0.0004\\
\end{tabular}
\end{table}

\begin{table}[h] 
\caption{Expansion (lab) energies $E_\mu$ (in MeV) used in the EST
representations of the quark-model potential.
$\epsilon_d$ refers to the deuteron binding energy.
$l_\mu$ is the boundary 
condition chosen for the angular momentum $l_\mu$ of the
initial state \protect\cite{NEMO,Haid84}.}
\begin{center} \begin{tabular}{lccccccccc}
Partial wave &\multicolumn{9}{c} {$(E_\mu,l_\mu)$} \\ \hline
$^1S_0(NN)-{^5D_0(N\Delta)}$  & (0,0) & (50,0) & (300,0) & (-20,0) & (-20,2)  &
(-50,0) & && \\
$^3S_1-{^3D_1}$ & $\epsilon_d$ &
(100,0) & (175,2) & (300,2) & (-50,0) & (-50,2) &&& \\
 \\ \end{tabular} \end{center}
\label{table3} 
\end{table}

\begin{table}[h]
\caption{Properties of the three-nucleon bound state.}
\label{table4}
\begin{tabular}{ccccc}
               &Quark model & 
Nijm II & Bonn B \protect\cite{WSH00}\\
\tableline 
$E_B$ (MeV)    & $-$7.72     &-7.65 & -8.17   \\ 
$P_S$ ($\%$)   & 91.49       &90.33&  91.35  \\ 
$P_{S'}$ ($\%$)& 1.430       &1.339&   1.368  \\ 
$P_P$ ($\%$)   & 0.044       &0.064&  0.049  \\ 
$P_D$ ($\%$)   & 7.033       &8.267&  7.235  \\ 
\end{tabular}
\end{table}

\newpage 

\begin{figure}[tbp]
\caption{$^1S_0$ $NN$ phase shift. The solid line is the result 
for the nonlocal quark-model potential. The dotted line shows the 
result of the separable representation of the nonlocal quark-model potential.
The squares, diamonds and triangles are the experimental data 
taken from 
Refs. \protect\cite{ex1}, \protect\cite{ex2}, and \protect\cite{ex3},
respectively.}
\label{fig1}
\end{figure}

\begin{figure}[tbp]
\caption{Phase shifts for the $^3S_1$ and $^3D_1$ partial waves. Same description as in Fig. \ref{fig1}.}
\label{fig2}
\end{figure}

\begin{figure}[tbp]
\caption{Mixing parameter $\varepsilon_1$. Same description as in Fig. \ref{fig1}.}
\label{fig3}
\end{figure}

\newpage 

\begin{figure}[hbct]
\epsfig{file=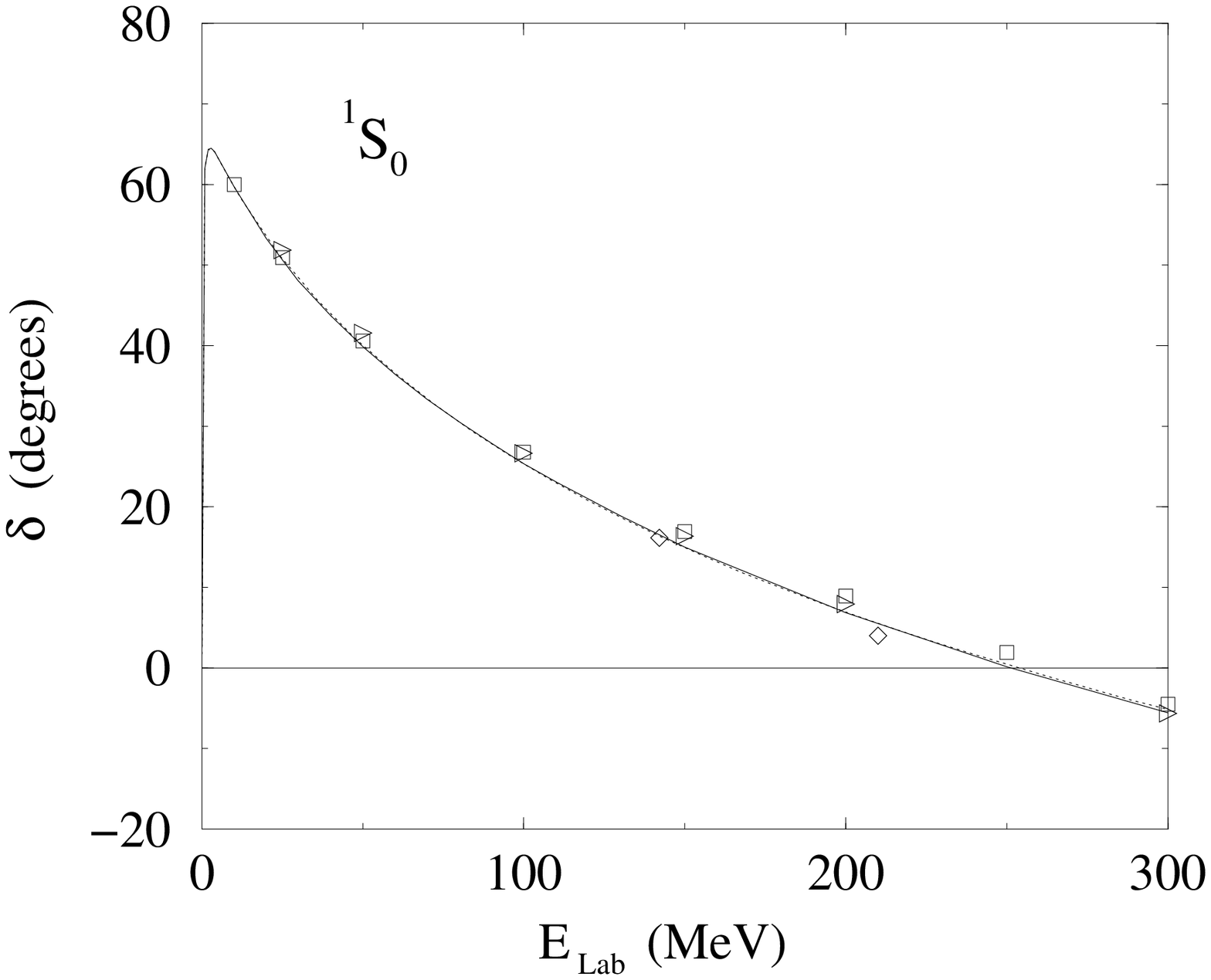, width=12cm}
\end{figure}
\vspace{5cm}
\center{Fig. 1}

\newpage 

\begin{figure}[hbct]
\epsfig{file=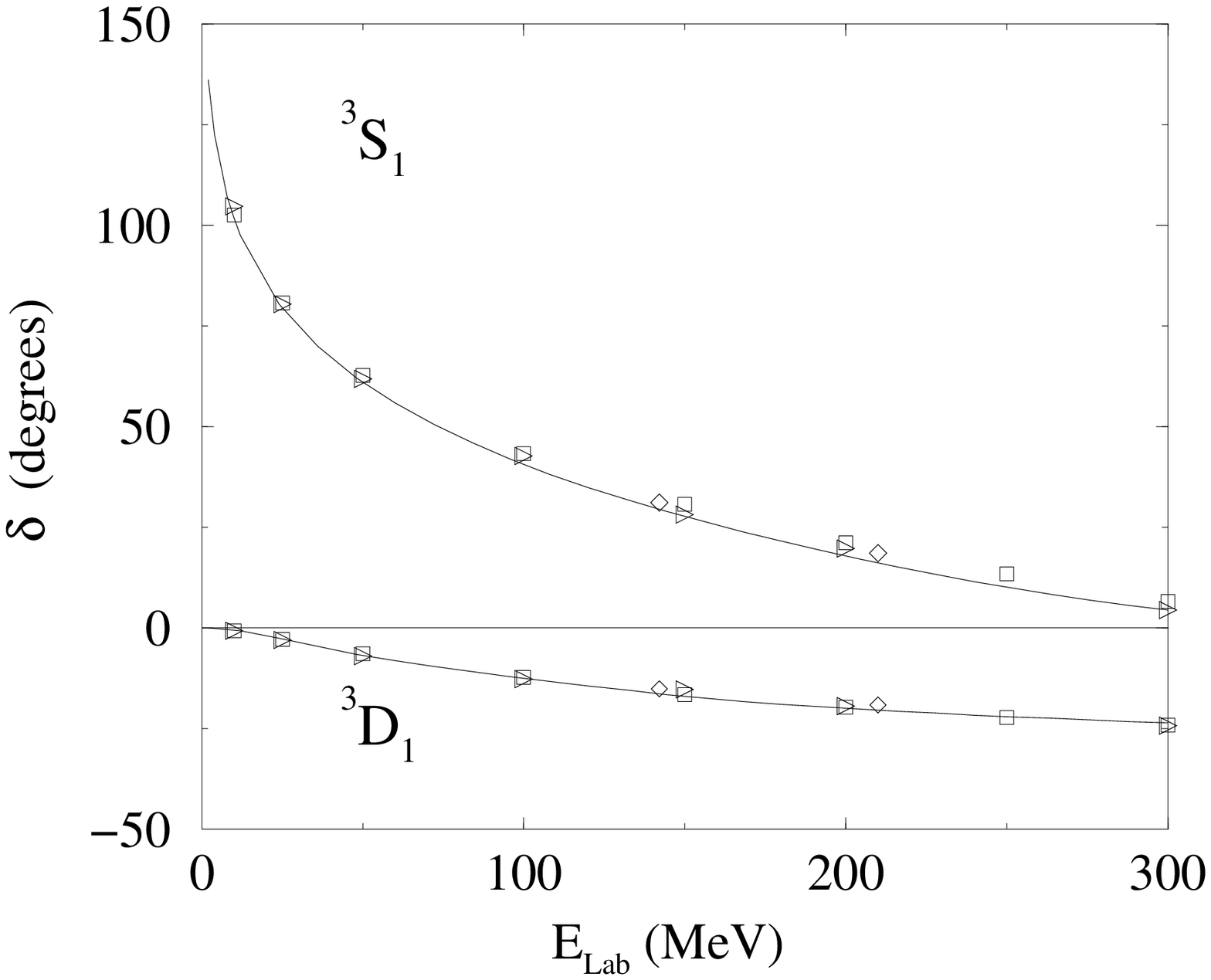, width=12cm}
\end{figure}
\vspace{5cm}
\center{Fig. 2}

\newpage 

\begin{figure}[hbct]
\epsfig{file=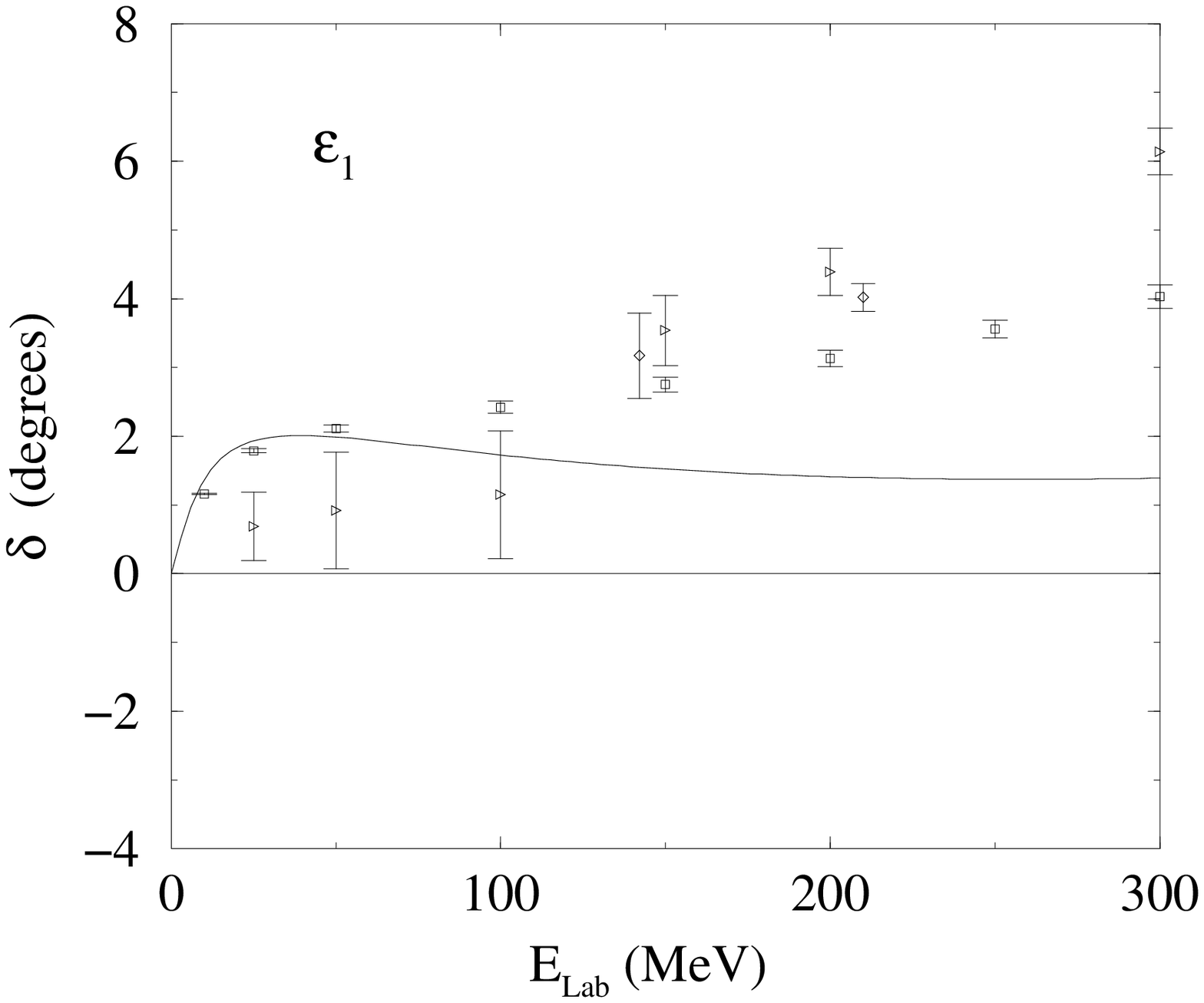, width=12cm}
\end{figure}
\vspace{5cm}
\center{Fig. 3}

\newpage

\end{document}